\documentstyle[aps,multicol,amsfonts,amssymb,amsmath,graphicx]{revtex}

\makeatletter
\def\thepage{\@arabic\c@page}
\def\@pnumwidth{2em}
\makeatother

\begin{document}

\def\text{\mbox}

\draft
\title{\Huge THE UNIVERSE'S EVOLUTION}
\author{N. T. Anh}
\address{Institute for Nuclear Science and Technique,\\
Hanoi, Vietnam. \\ {\footnotesize Email: anh@vaec.vista.gov.vn} }
\date{1999}
\maketitle

\begin{abstract}
Based on a new theory of causality [1] and its development to the
theory of the Universe [2], we show, in this paper, new ideas for
building a theory of everything.
\end{abstract}

\pacs{PACS No. : 01.55.+b}

\makeatletter
\global\@specialpagefalse
\def\@oddhead{N.T.Anh\hfill The Variant Principle}
\let\@evenhead\@oddhead
\def\@oddfoot{\reset@font\rm\hfill \thepage\hfill
\ifnum\c@page=1
  \llap{\protect\copyright{} 1999 N.T.Anh}%
\fi } \let\@evenfoot\@oddfoot
\makeatother

\begin{multicols}{2}
\section{Introduction}

\label{intro}The discovery of natural world around us is an indispensable
activity of mankind. And looking for a single theory that can explain every
phenomenon and every process is a good dream of scientists and especially
physicists. Nowadays, physicists have been trying to find a single theory
that unifies four familiar interactions, and they hope that it develops the
theory of everything. A theory they believe to be the theory of everything
is called the superstring theory. Notably, it is necessary to understand
that the superstring theory gives us description which only can unite four
familiar forces into a single framework.

Of course, whether it is really called the theory of everything or not since
there are many unknown interactions (besides four familiar interactions)
absenting in the theory. Moreover, the theory of everything must give us a
correct solution in every phenomenon and process (in all universe's
dimension, in all energy level, in all universe's status and so fourth). The
theory of everything is necessarily a theory of Creation, that is, it must
necessarily explain everything from the origin of the Universe down to the
lilies of the field. A theory of everything is also a theory of everyday.
Thus, this theory, when fully completed, will be able to explain the
existence of every phenomenon, the variation of every process, and many
others.

However, there is another way on which we can reach the theory of
everything. That is to find a single law that implies all known laws and,
therefore, predicts unknown laws. The presence of this law has in every
phenomenon, process and thing in nature. It is really to be the ultimate
goal of all knowledge, the theory to end all theories, the ultimate answer
to all questions.

The present article is the first one of a series that we would like to say
about the law and the theory of causality as well as its implementation for
building a theory of the Universe. We hope that some of the readers of this
article will find out that the law of causality is just the law of all laws,
the theory of causality is just the single theory of everything, and perhaps
they will be the ones to complete the quest for the Theory of the Universe.

The article is organized as follows. In Section 2 we introduce the ideas and
concepts for leading the equation of causality [1]. In Section 3, as the
main part of the article, we attempt to simulate briefly the process of the
Universe's evolution [2]. The conclusions and prospects are given in Section
4.

\section{The Equation of Causality}

We can always conceive that the Universe is in unification. And a surefire
fact is that the Universe's unification is only in a general intrinsic
relationship which is nothing but the relationship of causality, and the
unification only manifests itself in that causal form. Then, a question is
put on what is the ultimate cause of everything? On further reflection, we
find out that there exists an ultimate cause - that is the difference.

Truly, there would not exist anything if there were not the difference. If
there were no difference, this world did not exist. And there is a fact that
since the difference is the ultimate cause, it is the cause of itself, in
other words, it is also the effect of itself. The difference causes the
difference, the difference is the corollary generated by the difference.

In another way, we can imagine abstractly that the Nature is a set of
positive actions\footnote{%
Positive action means 'affirmation'.} and negative actions\footnote{%
Negative action means ''negative'.}. Then, what does the Nature act
positively on? and what does the Nature act negatively on? The answer to
these questions gives us a law. That is, what do not have any intrinsic
contradiction is acted positively on, what do have some intrinsic
contradiction is acted negatively on. Both the positive action (in front of
a process) and the negative action (in back of the process) have a final
goal that is to reach and to end at a new action.

Thus, we have started to come to a theory, axiom of which is the difference,
object of which is actions\footnote{%
Action here is a general concept of anything, it may be a function, a
generator, an operator, or even a force, an interaction, a field, ect.
depending on each considered subject.} [1].

Consider two actions, we obtain a definition that coexistence of two actions
which reject mutually generates contradiction. That is represented as
follows:
\[
M=\left\{
\begin{array}{c}
A\neq A\;\;\;\;\; -\text{Action }K_{1} \\
A=A\;\;\;\;\; -\text{Action }K_{2}
\end{array}
\text{.}\right.
\]

This means the higher the power of mutual rejection between two actions $%
K_{1}$and $K_{2}$ is the more severe the contradiction $M$ will be. And the
power of mutual rejection of two actions is estimated from the degree of
difference. A contradiction which is solved means that the difference of two
actions diminishes to zero. Herein, two actions $K_{1}$and $K_{2}$ all vary
to reach and to end at a new action $K_{3}$.

The change, and one kind of which - the variation, is generated by
contradiction. In exact words, the variation is the manifestation of
contradiction solving. The more severe contradiction becomes the more urgent
need of solving out of contradiction will be, and hence the more violent the
change, the variation of the state, i.e. of contradiction will become. Call
the violence, or the quickness of the variation of contradiction $Q$, the
contradiction state is $M$, the above principle can be represented as
follows:
\[
Q=K_{(M)}M
\]
where $K_{(M)}$ is means to solve the contradiction $M$. $K_{(M)}$ can be a
function of the contradiction state. It represents the degree of easiness to
escape the contradiction state. If the contradiction is characterized by
quantities $x,y,z,...$, these quantities themselves will be facilities to
transport the contradiction, degrees of freedom over which the contradiction
is solved. Hence, the degree of easiness is valued as the derivative of the
contradiction with respect to its degree of freedom
\[
K_{(M)}\thicksim \left| M^{\prime }(x,y,z,...)\right| .
\]
Thus,
\[
Q=a\left| M^{\prime }\right| M
\]
where the coefficient $a$ generates from choosing the dimension.

Advance a quantity $T$, inverse of $Q$, to be stagnancy of contradiction
solving. The sum of stagnancy in the process of contradiction solving from $%
M_{0}$ to $M_{0}-\Delta M$ we call the time is generated by this variation
\[
\Delta t\thickapprox -\frac{T+(T+\Delta T)}{2}\;\Delta M.
\]
Thus,
\[
\lim\nolimits_{\Delta T\rightarrow 0,\Delta M\rightarrow 0,\Delta
t\rightarrow 0}\frac{\Delta M}{\Delta t}=-\frac{1}{T}=-a\left|
M^{\prime }\right| M.
\]
Therefrom, we obtain the equation of causality,
\begin{equation}
\frac{dM}{dt}=-a\left| M^{\prime }(x,y,z,...)\right| M(x,y,z,...).
\end{equation}

Truthly, the difference is the origin of all, but it has the
meaning in direct relationship, in direct comparison. Some state
which has any intrinsic contradiction must vary to reach a new one
having no intrinsic contradiction, or exactly, \ having
infinitesimal contradiction. The greater the value of the
contradiction derivative with respect to some degree of freedom
is, the better the 'scent' for way out in that degree of freedom
will be, the greater the strength of the solved contradiction over
that degree of freedom will be.

It is easy to see that equation of causality (1) is represented as a
'classical' form. It can be developed to more general form in which the time
is considered as a new degree of freedom. However, Eq. (1) looks like
familiar equations, and we will use it for applying to concrete problems.
Though the law of causality (1) is abstract its concrete form in each
problem is very clear. And in the next Section we show the process of the
Universe's evolution which lays the foundation for building a theory of the
Universe.

\section{The Process of The Universe's Evolution}

\subsection{The general mechanism}

To survey clearly the evolution of the Universe, we firstly review four
important concepts: time, space, matter, and motion.

About the time. Can the time exist independently, if it is separated from
space, matter, and motion? Evidently, no. If the time were separated from
motion, the conception of it would have no meaning. The time cannot
self-exist, it is the effect of motion. No motion, no time.

About the motion. The motion also would not self-exist if it were separated
from matter and space.

And about the matter. The matter also cannot self-exist without space. It
exists owing to not only itself but also the coexistence of the space
surrounding it. In essence, the matter is nothing but just some space with
intrinsic relationship different from familiar space we see around us.

Imagine that all are vanished: matter, space,..., and in general, every
difference is vanished. Then, there exists only one. It is homogeneous and
limitless everywhere. It can self-exist. It is the first element. In this
unique there is nothing, but there exists the 'Nothing'. The Nothing is the
origin of all, the cause of all, since it has the first difference.

In Section 2, we have said the axiom of the theory of causality. That is the
difference. Imagine that if the present Universe has many differences, the
first state of the Universe will be the state which has fewest differences.
It is logical to show that the first Universe's state is the Nothing, and
the transformation chain ''difference - contradiction - solving'' is the
expansion of the Universe.

A remarkable consensus has been developing recently around what is called
''quantum cosmology'', which proposes a beautiful synthesis of seemingly
hostile viewpoints. In the beginning it was Nothing. No space, no matter or
energy. But according to the quantum principle, even Nothing was unstable.
Nothing began to decay, i.e. it began to 'boil', with billions of tiny
bubbles forming and expanding rapidly. Each bubble became an expanding
sub-universe\footnote{%
Our universe is actually part of a much larger ''multiverse'' of
sub-universes. Our sub-universe may co-exist with other sub-universes, but
our sub-universe may be one of the few compatible with life. This would
answer the age-old question of why the physics constants of the universe
fall in a narrow band compatible with the formation of life. If the
universal constants were changed slightly, then life would have been
impossible.}. Sub-universes can literally spring into existence as a quantum
fluctuation of Nothing. Resonances of vacuum fluctuations create first
elements of matter.

In Ref. [2] we show the elementary equation of Evolution
\begin{equation}
e^{\Sigma aM\,\widehat{\partial }}M=e^{\Sigma (-)\Delta t\,\stackrel{.}{%
\partial }}M,
\end{equation}
and the conservation relation of quanta
\begin{equation}
\sum\limits_{j,...,k}\sum\limits_{i=0}^{n}\frac{(-)^{i}}{n!}%
C_{i}^{n}M_{j}...M_{k}=0,
\end{equation}
where $n$ is total of quanta, $i$ is quantum number generated by each step
of expansion of the Universe, $C_{i}^{n}$ is binary coefficient.

It is easy to realize that Eq. (2) is also a form of equation of causality
(1). But Eq. (2) gives us an important application in modelling the
multiplication and the combination of quanta. There are two objects from Eq.
(2) we can use to study: one is actions, the other is quanta. Studying
actions gives us laws, equations, representations in each considered field.
And studying quanta gives us models, classifications, arrangements of
quanta. To describe the evolution of the Universe, it is better for us to
investigate quanta.

Call $\alpha ,\beta ,\gamma ,...$ quanta. For each quantum there is a rule
of multiplication as follows
\begin{equation}
\alpha ^{n}\rightarrow e^{\partial _{\alpha }}\alpha
^{n}=\sum_{i=0}^{n}C_{i}^{n}\alpha ^{n-i}=(\alpha +1)^{n}
\end{equation}
where $n$ is order of combination. Although Eq. (4) is obtained
from Eq. (2) in considering for quanta, it can be found meaningly
using the evolution principle shown in Ref. [2]. Eq. (4) itself
represents the evolution of the Universe.

\subsection{Examples for the doublet and the triplet}

Using Eq. (4) we consider two stages in the process of the Universe's
evolution: doublet and triplet.

For two interactive quanta the rule of multiplication reads
\end{multicols}
\rule{8.4cm}{.1mm}\rule{-.1mm}{.1mm}\rule{.1mm}{2mm}
\begin{equation}
\alpha ^{n},\beta ^{n}\rightarrow \frac{1}{2}(e^{\beta \partial _{\alpha
}}\alpha ^{n}+e^{\alpha \partial _{\beta }}\beta
^{n})=\sum_{i=0}^{n}C_{i}^{n}\alpha ^{n-i}\beta ^{i}=(\alpha +\beta )^{n}.
\end{equation}
And similar to three interactive quanta
\begin{eqnarray}
\alpha ^{n},\beta ^{n},\gamma ^{n}\rightarrow \frac{1}{3}(e^{(\beta +\gamma
)\partial _{\alpha }}\alpha ^{n}+e^{(\gamma +\alpha )\partial _{\beta
}}\beta ^{n}+e^{(\alpha +\beta )\partial _{\gamma }}\gamma ^{n})
&=&\sum_{m}^{n}\sum_{i}^{m}C_{m}^{n}C_{i}^{m}\alpha ^{n-m}\beta
^{m-i}\gamma ^{i}  \nonumber \\
&=&(\alpha +\beta +\gamma )^{n}.
\end{eqnarray}
And so fourth. Eqs. 5 and 6 can be drawn as schemata.

\begin{equation}
\begin{array}{cccccccccccccc}
\vdots &  &  &  & \cdots &  & \cdots &  & \cdots &  & \cdots &  &  &  \\
&  &  &  &  &  &  &  &  &  &  &  &  &  \\
\overline{2} &  &  &  &  &  & \overline{1} &  & \overline{1} &  &  &  &  &
\\
&  &  &  &  &  &  &  &  &  &  &  &  &  \\
0 &  &  &  &  &  &  & \bigcirc &  &  &  &  &  &  \\
&  &  &  &  &  &  &  &  &  &  &  &  &  \\
\underline{2} &  &  &  &  &  & 1 &  & 1 &  &  &  &  &  \\
&  &  &  &  &  &  &  &  &  &  &  &  &  \\
\underline{2}\otimes \underline{2}=\underline{3}\oplus \underline{1} &  &  &
&  & 1 &  & 2 &  & 1 &  &  &  &  \\
&  &  &  &  &  &  &  &  &  &  &  &  &  \\
\underline{2}\otimes \underline{2}\otimes \underline{2}=\underline{4}\oplus
\underline{2}\oplus \underline{2} &  &  &  & 1 &  & 3 &  & 3 &  & 1 &  &  &
\\
&  &  &  &  &  &  &  &  &  &  &  &  &  \\
\vdots &  &  & 1 &  & 4 &  & 6 &  & 4 &  & 1 &  &  \\
&  &  &  &  &  &  &  &  &  &  &  &  &  \\
\vdots &  & 1 &  & 5 &  & 10 &  & 10 &  & 5 &  & 1 &  \\
&  &  &  &  &  &  &  &  &  &  &  &  &  \\
\vdots & \cdots &  & \cdots &  & \cdots &  & \cdots &  & \cdots &  & \cdots
&  & \cdots
\end{array}
\end{equation}
\hspace{9cm}\rule{.1mm}{-4mm}\rule{.1mm}{.1mm}\rule{8cm}{.1mm}
\begin{multicols}{2}
is the schema for Eq. (5), where $\underline{2}$ means two quanta
$\alpha $ and $\beta $. The numbers in the triangle is the binary
coefficients which give us weights of classes. For example,
\[
\underline{2}\otimes \underline{2}=\underline{3}\oplus \underline{1}=
\begin{array}{ccccc}
&  & 1 &  &  \\
1 & \text{--------} & 1 & \text{--------} & 1
\end{array}
.
\]
And similar to Eq. (6) we have
\end{multicols}
\rule{8.4cm}{.1mm}\rule{-.1mm}{.1mm}\rule{.1mm}{2mm}
\begin{equation}
\begin{array}{cccccccccccccc}
\vdots &  &  &  & \overline{1} &  &  &  &  &  &  &  &  &  \\
\overline{3} &  &  & \overline{1} &  &  & \overline{1} &  &  &  &  &  &  &
\\
&  &  &  &  &  &  &  &  &  &  &  &  &  \\
&  &  &  &  &  &  &  &  &  &  &  &  &  \\
0 &  &  &  &  & \bigcirc &  &  &  &  &  &  &  &  \\
&  &  &  &  &  &  &  &  &  &  &  &  &  \\
&  &  &  &  &  &  &  &  &  &  &  &  &  \\
3 &  &  &  & {\bf 1} &  &  & {\bf 1} &  &  &  &  &  &  \\
&  &  &  &  &  & {\bf 1} &  &  &  &  &  &  &  \\
&  &  &  &  &  &  &  &  &  &  &  &  &  \\
&  &  &  &  &  &  &  &  &  &  &  &  &  \\
&  &  & {\bf 1} &  &  & 2 &  &  & {\bf 1} &  &  &  &  \\
\underline{3}\otimes \underline{3}=\underline{6}\oplus \overline{3} &  &  &
&  & {\bf 2} &  &  & {\bf 2} &  &  &  &  &  \\
&  &  &  &  &  &  & {\bf 1} &  &  &  &  &  &  \\
&  &  &  &  &  &  &  &  &  &  &  &  &  \\
&  & {\bf 1} &  &  & 3 &  &  & 3 &  &  & {\bf 1} &  &  \\
\underline{3}\otimes \underline{3}\otimes \underline{3}=\underline{10}\oplus
8\oplus 8\oplus 1 &  &  &  & {\bf 3} &  &  & 6 &  &  & {\bf 3} &  &  &  \\
&  &  &  &  &  & {\bf 3} &  &  & {\bf 3} &  &  &  &  \\
&  &  &  &  &  &  &  & {\bf 1} &  &  &  &  &  \\
&  &  &  &  &  &  &  &  &  &  &  &  &  \\
& {\bf 1} &  &  & 4 &  &  & 6 &  &  & 4 &  &  & {\bf 1} \\
&  &  & {\bf 4} &  &  & 12 &  &  & 12 &  &  & {\bf 4} &  \\
\underline{3}\otimes \underline{3}\otimes \underline{3}\otimes \underline{3}
&  &  &  &  & {\bf 6} &  &  & 12 &  &  & {\bf 6} &  &  \\
&  &  &  &  &  &  & {\bf 4} &  &  & {\bf 4} &  &  &  \\
\vdots &  &  &  &  &  &  &  &  & {\bf 1} &  &  &  &
\end{array}
\end{equation}
\hspace{9cm}\rule{.1mm}{-4mm}\rule{.1mm}{.1mm}\rule{8cm}{.1mm}
\begin{multicols}{2}
where $\underline{3}$ means three quanta $\alpha $, $\beta $ and
$\gamma $. The coefficients in the pyramid give us weights of
classes,
\[
\begin{tabular}{llllllll}
&  &  &  & $1$ &  &  &  \\
&  &  & ${\bf 1}$ &  &  & ${\bf 1}$ &  \\
$\underline{3}\otimes \underline{3}=\underline{6}\oplus \overline{3}=$ & $%
{\bf 1}$ &  &  & $1$ &  &  & ${\bf 1}$ \\
&  &  & ${\bf 1}$ &  &  & ${\bf 1}$ &  \\
&  &  &  &  & ${\bf 1}$ &  &
\end{tabular}
,
\]
\[
\begin{tabular}{llllllll}
&  &  &  & $1$ &  &  &  \\
$\underline{3}\otimes \overline{3}=1\oplus 8=$ &  & $1$ &  &  & $1$ &  &  \\
& ${\bf 1}$ &  &  & $2$ &  &  & ${\bf 1}$ \\
&  &  & ${\bf 1}$ &  &  & ${\bf 1}$ &
\end{tabular}
.
\]

It is easily to identify that the above schemata have the forms similar to
the $SU(2)$ and the $SU(3)$ groups. This means that for $n$ quanta we have a
corresponding schema according to the $SU(n)$ group, and the multiplication
and the combination of the Universe conform to the $SU$ group. And from
these schemata we can draw periodic diagrams of the Universe's quanta.

For simplification, we show below the periodic diagram of the two
quanta's multiplication made of the schema (7). Remodel (7) with
regard to the level splitting we have a new diagram,
\end{multicols}
\rule{8.4cm}{.1mm}\rule{-.1mm}{.1mm}\rule{.1mm}{2mm}
\[
\begin{tabular}{ccccccccc}
&  &  &  & \thinspace \fbox{{\tiny []}} &  &  &  &  \\
&  &  &  & {\tiny 1}\fbox{{\tiny ][}} &  &  &  &  \\
&  &  &  & {\tiny 2}\fbox{{\tiny ][}}\fbox{{\tiny ][}} &  &  &  &  \\
&  &  &  & {\tiny 5}\fbox{{\tiny ][}}\fbox{{\tiny ][}}\fbox{{\tiny ][}} &  &
&  &  \\
&  &  & {\tiny 1}\fbox{{\tiny ]}} & $\vdots $ & {\tiny 1}\fbox{{\tiny [}} &
&  &  \\
&  &  & {\tiny 2}\fbox{{\tiny ]}}\fbox{{\tiny ][}} &  & {\tiny 2}\fbox{%
{\tiny ][}}\fbox{{\tiny [}} &  &  &  \\
&  &  & {\tiny 5}\fbox{{\tiny ]}}\fbox{{\tiny ][}}\fbox{{\tiny ][}} &  &
{\tiny 5}\fbox{{\tiny ][}}\fbox{{\tiny ][}}\fbox{{\tiny [}} &  &  &  \\
&  &  & {\tiny 14}\fbox{{\tiny ]}}\fbox{{\tiny ][}}\fbox{{\tiny ][}}\fbox{%
{\tiny ][}} &  & {\tiny 14}\fbox{{\tiny ][}}\fbox{{\tiny ][}}\fbox{{\tiny ][}%
}\fbox{{\tiny [}} &  &  &  \\
&  & {\tiny 1}\fbox{{\tiny ]}}\fbox{{\tiny ]}} & $\vdots $ & $\backslash$\qquad {\tiny 1}\fbox{{\tiny ][}}\qquad / & $\vdots $ & {\tiny 1}%
\fbox{{\tiny [}}\fbox{{\tiny [}} &  &  \\
&  & {\tiny 3}\fbox{{\tiny ]}}\fbox{{\tiny ]}}\fbox{{\tiny ][}} &  & {\tiny 3%
}\fbox{{\tiny ][}}\fbox{{\tiny ][}} &  & {\tiny 3}\fbox{{\tiny ][}}\fbox{%
{\tiny [}}\fbox{{\tiny [}} &  &  \\
&  & {\tiny 9}\fbox{{\tiny ]}}\fbox{{\tiny ]}}\fbox{{\tiny ][}}\fbox{{\tiny %
][}} &  & {\tiny 9}\fbox{{\tiny ][}}\fbox{{\tiny ][}}\fbox{{\tiny ][}} &  &
{\tiny 9}\fbox{{\tiny ][}}\fbox{{\tiny ][}}\fbox{{\tiny [}}\fbox{{\tiny [}}
&  &  \\
&  & $\vdots $ & \multicolumn{1}{l}{$\backslash$} & {\tiny 28}%
\fbox{{\tiny ][}}\fbox{{\tiny ][}}\fbox{{\tiny ][}}\fbox{{\tiny ][}} &
\multicolumn{1}{r}{/} & $\vdots $ &  &  \\
& {\tiny 1}\fbox{{\tiny ]}}\fbox{{\tiny ]}}\fbox{{\tiny ]}} &  & {\tiny 1}%
\fbox{{\tiny ]}}\fbox{{\tiny ][}} & $\vdots $ & {\tiny 1}\fbox{{\tiny ][}}%
\fbox{{\tiny [}} &  & {\tiny 1}\fbox{{\tiny [}}\fbox{{\tiny [}}\fbox{{\tiny [%
}} &  \\
& {\tiny 4}\fbox{{\tiny ]}}\fbox{{\tiny ]}}\fbox{{\tiny ]}}\fbox{{\tiny ][}}
&  & {\tiny 4}\fbox{{\tiny ]}}\fbox{{\tiny ][}}\fbox{{\tiny ][}} &  & {\tiny %
4}\fbox{{\tiny ][}}\fbox{{\tiny ][}}\fbox{{\tiny [}} &  & {\tiny 4}\fbox{%
{\tiny ][}}\fbox{{\tiny [}}\fbox{{\tiny [}}\fbox{{\tiny [}} &  \\
& $\vdots $ & \multicolumn{1}{l}{$\backslash$} & {\tiny 14}\fbox{%
{\tiny ]}}\fbox{{\tiny ][}}\fbox{{\tiny ][}}\fbox{{\tiny ][}} &  & {\tiny 14}%
\fbox{{\tiny ][}}\fbox{{\tiny ][}}\fbox{{\tiny ][}}\fbox{{\tiny [}} &
\multicolumn{1}{r}{/} & $\vdots $ &  \\
{\tiny ...} &  & {\tiny 1}\fbox{{\tiny ]}}\fbox{{\tiny ]}}\fbox{{\tiny ][}}
& $\vdots $ & $\backslash$\quad {\tiny 1}\fbox{{\tiny ][}}\fbox{%
{\tiny ][}}\quad / & $\vdots $ & {\tiny 1}\fbox{{\tiny ][}}\fbox{{\tiny [}}%
\fbox{{\tiny [}} &  & {\tiny ...} \\
{\tiny ...} &  & {\tiny 5}\fbox{{\tiny ]}}\fbox{{\tiny ]}}\fbox{{\tiny ][}}%
\fbox{{\tiny ][}} &  & {\tiny 5}\fbox{{\tiny ][}}\fbox{{\tiny ][}}\fbox{%
{\tiny ][}} &  & {\tiny 5}\fbox{{\tiny ][}}\fbox{{\tiny ][}}\fbox{{\tiny [}}%
\fbox{{\tiny [}} &  & {\tiny ...} \\
&  & $\vdots $ & \multicolumn{1}{l}{$\backslash$} & {\tiny 20}%
\fbox{{\tiny ][}}\fbox{{\tiny ][}}\fbox{{\tiny ][}}\fbox{{\tiny ][}} &
\multicolumn{1}{r}{/} & $\vdots $ &  &  \\
{\tiny ...} & {\tiny 1}\fbox{{\tiny ]}}\fbox{{\tiny ]}}\fbox{{\tiny ]}}\fbox{%
{\tiny ][}} &  & {\tiny 1}\fbox{{\tiny ]}}\fbox{{\tiny ][}}\fbox{{\tiny ][}}
& $\vdots $ & {\tiny 1}\fbox{{\tiny ][}}\fbox{{\tiny ][}}\fbox{{\tiny [}} &
& {\tiny 1}\fbox{{\tiny ][}}\fbox{{\tiny [}}\fbox{{\tiny [}}\fbox{{\tiny [}}
& {\tiny ...} \\
& $\vdots $ &  & {\tiny 6}\fbox{{\tiny ]}}\fbox{{\tiny ][}}\fbox{{\tiny ][}}%
\fbox{{\tiny ][}} &  & {\tiny 6}\fbox{{\tiny ][}}\fbox{{\tiny ][}}\fbox{%
{\tiny ][}}\fbox{{\tiny [}} &  & $\vdots $ &  \\
{\tiny ...} &  & {\tiny 1}\fbox{{\tiny ]}}\fbox{{\tiny ]}}\fbox{{\tiny ][}}%
\fbox{{\tiny ][}} & $\vdots $ & $\backslash$\quad {\tiny 1}\fbox{%
{\tiny ][}}\fbox{{\tiny ][}}\fbox{{\tiny ][}}\quad / & $\vdots $ & {\tiny 1}%
\fbox{{\tiny ][}}\fbox{{\tiny ][}}\fbox{{\tiny [}}\fbox{{\tiny [}} &  &
{\tiny ...} \\
&  & $\vdots $ &  & {\tiny 7}\fbox{{\tiny ][}}\fbox{{\tiny ][}}\fbox{{\tiny %
][}}\fbox{{\tiny ][}} &  & $\vdots $ &  &  \\
&  &  &  & $\vdots $ &  &  &  &
\end{tabular}
\]

Arrange this diagram in the order of the levels we obtain the so-called
periodic diagram

\[
\begin{tabular}{p{0.01cm}p{0.01cm}p{0.01cm}p{0.01cm}p{0.01cm}p{0.01cm}p{0.01cm}p{0.01cm}p{0.01cm}p{0.01cm}p{0.01cm}p{0.01cm}p{0.01cm}p{0.01cm}p{0.01cm}p{0.01cm}p{0.01cm}p{0.01cm}p{0.01cm}p{0.01cm}p{0.01cm}}
&  &  &  &  &  &  &  &  & $\bullet $ &  &  &  &  &  &  &  &  &  &  &
\multicolumn{1}{|p{0.01cm}}{I} \\ \cline{1-20}
&  &  &  &  &  &  &  &  & $\bullet $ &  &  &  &  &  &  &  &  &  &  &  \\
&  &  &  &  &  &  &  & $\diagup $ &  & $\diagdown $ &  &  &  &  &  &  &  &
&  &  \\
&  &  &  &  &  &  & $\bullet $ & $\leftharpoonup $ & $\bullet $ & $%
\rightharpoonup $ & $\bullet $ &  &  &  &  &  &  &  &  &
\multicolumn{1}{|p{0.01cm}}{II} \\ \cline{1-20}\cline{20-20}
&  &  &  &  &  &  &  &  & $\bullet $ &  &  &  &  &  &  &  &  &  &  &  \\
&  &  &  &  &  &  &  & $\diagup $ &  & $\diagdown $ &  &  &  &  &  &  &  &
&  &  \\
&  &  &  &  &  &  & $\bullet $ & $-$ & $\bullet $ & $-$ & $\bullet $ &  &  &
&  &  &  &  &  & \multicolumn{1}{|p{0.01cm}}{III} \\ \cline{1-20}
&  &  &  &  &  & $\diagup $ &  &  & $\bullet $ &  &  & $\diagdown $ &  &  &
&  &  &  &  &  \\
&  &  &  &  & $\bullet $ & $\leftharpoonup $ & $\bullet $ & $\diagup $ & $%
\bullet $ & $\diagdown $ & $\bullet $ & $\rightharpoonup $ & $\bullet $ &  &
&  &  &  &  &  \\
&  &  &  &  &  &  & $\bullet $ & $-$ & $\bullet $ & $-$ & $\bullet $ &  &  &
&  &  &  &  &  & \multicolumn{1}{|p{0.01cm}}{IV} \\ \cline{1-20}
&  &  &  &  &  & $\diagup $ &  &  & $\bullet $ &  &  & $\diagdown $ &  &  &
&  &  &  &  &  \\
&  &  &  &  & $\bullet $ & $-$ & $\bullet $ & $\diagup $ & $\bullet $ & $%
\diagdown $ & $\bullet $ & $-$ & $\bullet $ &  &  &  &  &  &  &  \\
&  &  &  & $\diagup $ &  &  & $\bullet $ & $-$ & $\bullet $ & $-$ & $\bullet
$ &  &  & $\diagdown $ &  &  &  &  &  & \multicolumn{1}{|p{0.01cm}}{V} \\
\cline{1-20}
&  &  & $\diagup $ &  &  & $\diagup $ &  &  & $\bullet $ &  &  & $\diagdown $
&  &  & $\diagdown $ &  &  &  &  &  \\
&  & $\bullet $ & $\leftharpoonup $ & $\bullet $ & $\diagup $ & $\bullet $ &
$-$ & $\diagup $ & $\bullet $ & $\diagdown $ & $-$ & $\bullet $ & $\diagdown
$ & $\bullet $ & $\rightharpoonup $ & $\bullet $ &  &  &  &  \\
&  &  &  & $\bullet $ & $-$ & $\bullet $ & $\diagup $ & $-$ & $\bullet $ & $%
- $ & $\diagdown $ & $\bullet $ & $-$ & $\bullet $ &  &  &  &  &  &  \\
&  &  & $\diagup $ &  &  & $\bullet $ & $-$ & $-$ & $\bullet $ & $-$ & $-$ &
$\bullet $ &  &  & $\diagdown $ &  &  &  &  & \multicolumn{1}{|p{0.01cm}}{VI}
\\ \cline{1-20}
&  & $\diagup $ &  &  & $\diagup $ &  &  &  & $\bullet $ &  &  &  & $%
\diagdown $ &  &  & $\diagdown $ &  &  &  &  \\
& $\bullet $ & $\leftharpoonup $ & $\bullet $ & $\diagup $ & $-$ & $\bullet $
& $-$ & $\diagup $ & $\bullet $ & $\diagdown $ & $-$ & $\bullet $ & $-$ & $%
\diagdown $ & $\bullet $ & $\rightharpoonup $ & $\bullet $ &  &  &  \\
&  &  & $\bullet $ & $-$ & $-$ & $\bullet $ & $\diagup $ & $-$ & $\bullet $
& $-$ & $\diagdown $ & $\bullet $ & $-$ & $-$ & $\bullet $ &  &  &  &  &  \\
&  & $\diagup $ &  &  &  & $\bullet $ & $-$ & $-$ & $\bullet $ & $-$ & $-$ &
$\bullet $ &  &  &  & $\diagdown $ &  &  &  & \multicolumn{1}{|p{0.01cm}}{VII
} \\ \cline{1-20}
& $\diagup $ &  &  &  & $\diagup $ &  &  &  & $\bullet $ &  &  &  & $%
\diagdown $ &  &  &  & $\diagdown $ &  &  &  \\
$\bullet $ & $\leftharpoonup $ & $-$ & $\bullet $ & $\diagup $ & $\bullet $
& $-$ & $\bullet $ & $\diagup $ & $\bullet $ & $\diagdown $ & $\bullet $ & $%
- $ & $\bullet $ & $\diagdown $ & $\bullet $ & $-$ & $\rightharpoonup $ & $%
\bullet $ &  &  \\
&  &  & $\diagup $ &  &  &  & $\diagup $ &  &  &  & $\diagdown $ &  &  &  & $%
\diagdown $ &  &  &  &  &
\end{tabular}
,
\]
which is nothing but the Mendeleev periodic table built in the energy levels,

\begin{figure}[h]
\begin{center}
\leavevmode
\includegraphics[width=0.5\columnwidth]{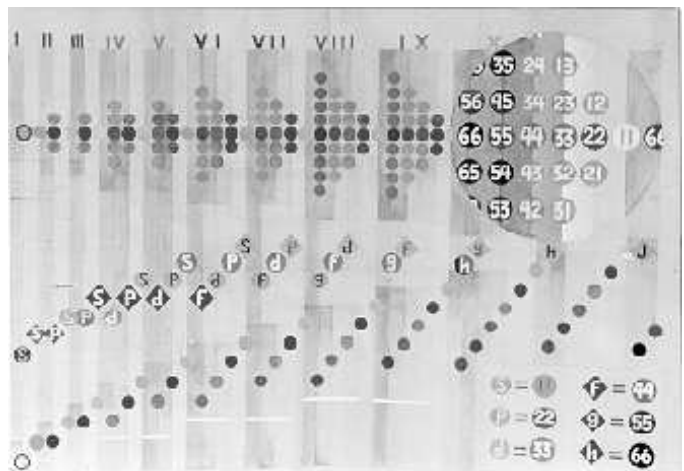}
\caption{The pine-tree form of the periodic law}
\end{center}
\end{figure}
and we can paint it as an abstract picture,
\begin{figure}[h]
\begin{center}
\includegraphics[width=0.5\columnwidth]{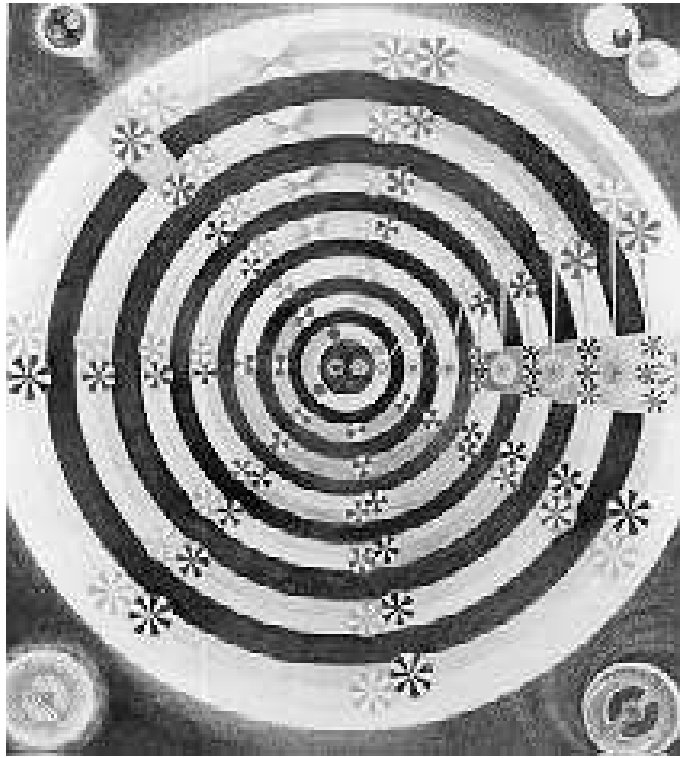}
\caption{The spiral form of the periodic law}
\end{center}
\end{figure}
\hspace{9cm}\rule{.1mm}{-4mm}\rule{.1mm}{.1mm}\rule{8cm}{.1mm}
\begin{multicols}{2}
The pictures Fig.1 and Fig.2 have a very special significance
besides the periodic law. Thus, corresponding to the $SU(2)$,
$SU(3)$, and $SU(n)$ groups we have the periodic laws of doublet,
triplet, and multiplet elements. They give us a model of the
evolution in the pine-tree and the spiral from simplex to complex,
from low-level to high-level.

\section{Conclusions and Prospects}

There is a truth that everybody knows: the nature is difficult to understand
for us when it has not been discovered yet, but it is really beautiful when
we understand it. This is science, where the ultimate worth of one's ideas
is that they lead to a genuine understanding of nature. And an idea or a
theory not only represents daily phenomena but also makes predictions that
survive comparison with observation and experiment based on fundamental
principles and laws that underlie the universe. By the present article, we
can confirm an existence of an ultimate principle or an ultimate law from
which others could be found out.

We realize that the most important principle of nature is that all
observable properties of things are about relationships. The difference has
meaning in direct relationship. Actions are in interaction in mutual
relationships. Contradictions are generated in mutual-rejection
relationships. Transformation, change, or motion, variation, or exactly
contradiction solving, does experience of relationships. Even space and time
must be spoken about in terms of relationships. There is no such a thing as
space independent of that which exists in it and no such thing as time apart
from change. These mean that the universe is in unification, and this
unification is created by relationships of causality.

Relationships of causality give us an ultimate law which is called the law
of causality. Following the logical source of the law of causality, we open
up limitless horizons of a view of the universe. The Universe was born from
Nothing, and its evolution created beautiful worlds of numerous form of
things whose structure and complexity can be self-organized. We understand
that there are natural processes, easily comprehensible, by which
organization can arise naturally and spontaneously, without any need for a
maker outside of the system. That is confirmed in the present article.

Although the results we obtained in this article is similar to ones that
modern physics discovered, we open to the possibility that the answers to
many of the questions we have about why phenomena, things, the elementary
particles, or the fundamental forces are as they are and not otherwise, and
why the nature created beautiful worlds in the way we see not otherwise.
Moreover, we have the expectation to answer the greatest questions: ''Where
does the universe come from?'' or ''What is the evolution, the
self-organization, the variety, or the fate of the universe?'' or ''Where
does the matter come from and where is the missing matter?''.

In the present article's view of the universe, everything is from to
nothing, everything may be smooth at the beginning but does not stay smooth
forever, because today our universe is very inhomogeneous. So the universe
was not perfectly homogeneous either when it began or shortly after it began
but, rather, it was slightly inhomogeneous. It had small regions where the
density of matter was slightly higher than average and other regions where
it was slightly lower than average. They are really tiny. Yet tiny as they
are to begin with, these inhomogeneities are very important because they are
the seeds from which particles, star clusters, galaxies and, eventually,
human beings, will grow in the way that their structure must be formed
systematically from within by natural processes of self-organization such as
periodic, multiplicative, combinative, evolutive, and etc. principles.

Our universe has a variety of mysteries to discover. But we cannot say
everything in a day. Many and very many beautiful worlds are in future of
our discovery. This article is only the first one we would like to open up a
first view of the universe. The first is the key idea behind evolution of
the universe from nothing, the second the idea behind the principle of
causality. These themes are only essential for understanding what happened,
is happening, and will happen in the universe.

Of course, this does not mean that theories will be discovered, based on the
principle of causality, are proven to be right; only observation and
experiment can, in the end, tell us that. But a definite fact that we enter
the 21st century with new ideas and wide horizons, with much to do and
everything to talk about.

\section*{Acknowledgments}

We would like to thank Dr. D. M. Chi for useful discussions and valuable
comments.

The present article was supported in part by the Advanced Research Project
on Natural Sciences of the MT\&A Center.

\end{multicols}
\end{document}